\documentclass[ALICE,manyauthors]{cernphprep}

\usepackage{graphicx, subfigure}
\usepackage{amssymb}
\usepackage{amsmath}
\usepackage{lineno}
\usepackage{cite}
\usepackage{mciteplus}
\usepackage{ulem} 
\newcommand{\pt}{\ensuremath{p_\mathrm T}}
\usepackage{multirow}

\begin{document}%
%
%
\begin{titlepage}
\PHnumber{2012-233}                 
\PHdate{12 Aug 2012}              
%
%
\title{Centrality dependence of charged particle production at large transverse momentum in \mbox{Pb--Pb} collisions at $\sqrt{s_{\rm{NN}}} = 2.76$ TeV}
\ShortTitle{Particle production at large transverse momentum}   
%
\Collaboration{ALICE Collaboration%
         \thanks{See Appendix~\ref{app:collab} for the list of collaboration
                      members}}
\ShortAuthor{ALICE Collaboration}      

\begin{abstract}

The inclusive transverse momentum ($p_{\rm T}$) distributions of
primary charged particles are measured in the pseudo-rapidity range
$|\eta|<0.8$ as a function of event centrality in \mbox{Pb--Pb} collisions  at $\sqrt{s_{\rm{NN}}}=2.76$~TeV with ALICE at the LHC. 
The data are presented in the $p_{\rm T}$ range \mbox{$0.15<p_{\rm
T}<50$~GeV/$c$} for nine centrality intervals from \mbox{70--80\%} to
\mbox{0--5\%}. 
The results in \mbox{Pb--Pb} are presented in terms of the nuclear
modification factor $R_{\rm{AA}}$ using a pp reference spectrum measured at the same collision energy.
We observe that the suppression of high-$p_{\rm T}$ particles strongly depends on event
centrality. The yield is most suppressed in central collisions (0--5\%) with $R_{\rm{AA}}\approx0.13$ at \mbox{$p_{\rm T}=6$--7~GeV/$c$}.
Above $p_{\rm T}=7$~GeV/$c$, there is a
significant rise in the nuclear modification factor, which reaches
$R_{\rm{AA}} \approx0.4$ for $p_{\rm T}>30$~GeV/$c$. In peripheral
collisions (70--80\%), only moderate suppression ($R_{\rm{AA}}=0.6$--0.7) and a weak $\pt$ dependence is observed.
The measured nuclear modification factors
are compared to other measurements and model calculations.

\end{abstract}
\end{titlepage}
\setcounter{page}{2}
\section{Introduction}

High-energy collisions of heavy-ions enable the study of hot
and dense strongly interacting matter \cite{BRAHMS_WHITE,PHOBOS_WHITE,STAR_WHITE,PHENIX_WHITE,SPS}. At
sufficiently high temperature, it is expected that partons (quarks and gluons)
are the dominant degrees of freedom. During the very early stage of
the collision, some of the incoming partons experience scatterings
with large momentum
transfers. These partons lose energy when they traverse the hot and dense medium that is formed.
One of the major goals of the heavy-ion physics programme at the LHC is to understand the underlying mechanisms for parton energy loss and use this as a tool to probe the properties of the medium.

Parton energy loss in heavy-ion collisions was first observed at RHIC as the suppression of high-$p_{\rm T}$ particle production in \mbox{Au--Au} collisions compared to expectations from an independent superposition of nucleon-nucleon collisions  \cite{PHENIXRAA130, STARRAA130, STARRAA, PHENIXRAA}. At RHIC, the particle production in central (0-5\%) \mbox{\mbox{Au--Au}} collisions at $\sqrt{s_{\rm{NN}}}=200$ GeV is suppressed by a factor of 5 at \mbox{$p_{\rm T}=5$--6~GeV/$c$} \cite{STARRAA, PHENIXRAA}, and is consistent with being independent of $\pt$ over the measured range $5<p_{\rm T}<20$~GeV/$c$ \cite{PHENIXPI0RAA}.

The increase of the charged particle density (${\rm{d}}N_{\rm{ch}}/\rm{d}{\eta}$) at mid-rapidity from RHIC energies to actual LHC energies by
a factor of around 2.2 \cite{ALICE_MULT} implies a similar increase in energy
density. However, the observed suppression of high-$\pt$ particle
production also depends on the ratio of quarks to gluons due to
their different color factors, and on the steepness of the $\pt$ spectra of the scattered partons. At the LHC the initial parton $p_{\rm T}$ spectra are less
steep than at RHIC and the ratio of gluons to quarks at a
given $p_{\rm T}$ is higher \cite{Eskola:2004cr}. The measurement of
high-$\pt$ hadron production at  the LHC helps to
disentangle the effects which cause the suppression and
provides a critical test of existing energy loss calculations
\cite{MODELS}. In particular, the large $p_{\rm T}$ reach provides a means to study the dependence of the energy loss on the initial parton energy.

 We present a measurement of the  $p_{\rm T}$ distributions of
charged particles in $0.15<p_{\rm T}<50$~GeV/$c$ with pseudo-rapidity $|\eta| < 0.8$, where \begin{math} \eta = -{\rm{ln}}[{\rm{tan}} (\theta/2)]
\end{math}, with $\theta$ the polar angle between the charged particle
direction and the beam axis. Results are presented for different centrality
intervals in \mbox{Pb--Pb} collisions at $\sqrt{s_{\rm{NN}}}=2.76$~TeV. They
are compared with measurements in pp collisions, by calculating the
nuclear modification factor

\begin{equation}\label{EQUATION_RAA}
R_{\rm{AA}} (p_{\rm T} ) = \frac{{\rm d}^{2}N^{\rm{AA}}_{\rm{ch}}/{\rm d}\eta{\rm d}p_{\rm{T}} }{\langle T_{\rm {AA}} \rangle {\rm d}^{2}\sigma^{\rm{pp}}_{ch}/{\rm d}\eta {\rm d}p_{\rm T} }
\end{equation}
where $N^{\rm{AA}}_{\rm{ch}}$ and $\sigma^{\rm{pp}}_{\rm{ch}}$ represent the charged particle yield in nucleus-nucleus (AA) collisions and the cross section in pp collisions, respectively. The nuclear
overlap function $T_{\rm{AA}}$ is calculated from the 
Glauber model \cite{MILLER_TAA} and averaged over each centrality interval, $\langle T_{\rm{AA}}  \rangle = \langle N_{\rm{coll}}  \rangle / \sigma^{\rm{NN}}_{\rm{inel}}$, where $\langle N_{\rm{coll}} \rangle$ is the average number of binary nucleon-nucleon collisions and $\sigma^{\rm{NN}}_{\rm{inel}}$ is the inelastic nucleon-nucleon cross section.

Early results from ALICE \cite{ALICE_RAA} showed that the production of charged
particles in central (0--5\%) \mbox{Pb--Pb} collisions at
$\sqrt{s_{\rm{NN}}} = 2.76$ TeV is suppressed by more than a factor of $6$ at $p_{\rm
 T}=6$--7~GeV/$c$ compared to an independent superposition of
nucleon-nucleon collisions, and that the suppression is stronger than that observed at RHIC.  The present data extend the study of high-$p_{\rm T}$ particle suppression in \mbox{Pb--Pb} out to $p_{\rm T}=50$ GeV/$c$ with a systematic study of the centrality dependence.

Moreover, the systematic uncertainties related to the pp reference were significantly reduced with respect to the previous measurement by using the $p_{\rm T}$ distribution measured in pp collisions at $\sqrt{s}=2.76$ TeV \cite{PPREFERENCE}. 

\begin{table}
\centering
\caption{\label{CENTR} Average values of the number of
 participating nucleons $\langle N_{{\rm{part}}} \rangle$
and the nuclear overlap function $\langle T_{\rm{AA}} \rangle$ \cite{MILLER_TAA} for the centrality intervals used in the analysis.}
\begin{tabular}{|l|l|l|l|}
\hline
Centrality&$\langle N_{{\rm{part}}}\rangle$ 
&$\langle T_{{\rm{AA}}} \rangle$ (mb$^{-1}$)\\
\hline
0--5\% & 383  $\pm$ 3  & 26.4 $\pm$ 1.1 \\
5--10\% & 330 $\pm$ 5  & 20.6 $\pm$ 0.9 \\
10--20\% & 261 $\pm$ 4   & 14.4 $\pm$ 0.6 \\
20--30\% & 186  $\pm$ 4   & 8.7 $\pm$ 0.4 \\
30--40\% & 129  $\pm$ 3   & 5.0 $\pm$ 0.2 \\
40--50\% & 85   $\pm$ 3  & 2.68 $\pm$ 0.14 \\
50--60\% & 53 $\pm$ 2  & 1.32 $\pm$ 0.09 \\
60--70\% & 30.0 $\pm$ 1.3  & 0.59 $\pm$ 0.04 \\
70--80\% & 15.8 $\pm$ 0.6  & 0.24 $\pm$ 0.03 \\
\hline
\end{tabular}
\end{table}

\section{Experiment and Data Analysis}

The ALICE detector is described in \cite{ALICE_DET}.  The Inner Tracking System (ITS) and the Time Projection Chamber (TPC) are used for vertex finding and tracking. The minimum-bias interaction trigger was derived from signals from the forward scintillators (VZERO), and the two innermost layers of the ITS (Silicon Pixel Detector - SPD). The collision centrality is determined using the VZERO. In addition, the information from two neutron Zero Degree Calorimeters (ZDCs) positioned at $\pm114$~m from the interaction point was used to remove contributions from beam-gas and electromagnetic interactions. The trigger and centrality selection are described in more detail in \cite{ALICE_MULT}.

The following analysis is based on $1.6\cdot10^{7}$ minimum-bias
\mbox{Pb--Pb} events recorded by ALICE in 2010. For this study, the events
are divided into nine centrality intervals from the 70--80\% to the 0--5\% most central \mbox{Pb--Pb} collisions, expressed in
percentage of the total hadronic cross section. The event centrality
can be related to the number of participating nucleons
$N_{{\rm{part}}}$ and the nuclear overlap function $T_{{\rm{AA}}}$ by using
simulations based on the Glauber model \cite{MILLER_TAA}. The average
values of $N_{{\rm{part}}}$ and $T_{\rm{AA}}$ for each centrality
interval,  $\langle N_{{\rm{part}}} \rangle$ and $\langle T_{\rm{AA}}  \rangle$, along with their corresponding systematic uncertainties, are
listed in \mbox{Table~\ref{CENTR}}. The errors include the experimental
uncertainties on the inelastic nucleon-nucleon cross section
\mbox{$\sigma^{\rm{NN}}_{\rm{inel}}=64 \pm 5$~mb} at $\sqrt{s_{\rm{NN}}}=2.76$~TeV \cite{CROSSSEC} and on the parameters of the nuclear
density profile used in the Glauber simulations  (more details in \cite{ALICE_MULT}).


The primary vertex position was determined from the tracks reconstructed in the ITS and the TPC by using an analytic $\chi^2$ minimization method, applied after approximating each of the tracks by a straight line in the vicinity of their common origin. The event is accepted if the
coordinate of the reconstructed vertex measured along the beam direction
($z$-axis) is within \mbox{$\pm10$~cm} around the nominal interaction point. The event vertex reconstruction is fully efficient
for the event centralities covered.

Primary charged particles are defined as all prompt particles produced
in the collision, including decay products, except those from weak
decays of strange hadrons. A set of standard cuts based on the number
of space points and the quality of the momentum fit in the TPC and ITS is
applied to the reconstructed tracks. Track candidates in the TPC are required to have hits in at least 120 (out of a maximum of 159) pad-rows and $\chi^{2}$ per point of the momentum fit smaller than 4.  Such tracks are projected to the ITS and used for further analysis if at least 2 matching hits (out of a maximum of 6) in the ITS, including at least one in the SPD,  are found. In addition,  the $\chi^{2}$ per point of the momentum fit in the ITS must be smaller than 36. In order to improve the purity of primary track
reconstruction at high $p_{\rm T}$ we developed a procedure where we
compare tracking information from the combined ITS and TPC track
reconstruction algorithm to that derived only from the TPC and
constrained by the interaction vertex point. We calculated the $\chi^{2}_{{\rm{TPC-ITS}}}$ between these tracks using the following formula

\begin{align} \label{CHI2FORMULA}
\chi^{2}_{{\rm{TPC-ITS}}} = & ({\bf{v_{TPC}}}-{\bf{v_{TPC-ITS}}})^{\bf{T}}  \cdot ({\bf{C_{TPC}}}+{\bf{C_{TPC-ITS}}})^{\bf{-1}} \cdot ({\bf{v_{TPC}}}-{\bf{v_{TPC-ITS}}})
\end{align}
 where $\bf{v_{TPC}}$, $\bf{v_{TPC-ITS}}$ and $\bf{C_{TPC}}$, $\bf{C_{TPC-ITS}}$ represent the measured track parameter vectors \\ ${\bf v}=(x,y,z,\theta,\phi,1/\pt)$ and their covariance matrices, respectively.   
If the $\chi_{\rm{TPC-ITS}}^{2}$ is larger than 36 the
track candidate is rejected. At $p_{\rm T}=0.15$--50~GeV/$c$, this procedure removes about 2--7\% (1--3\%) of the reconstructed tracks in the most central (peripheral) collisions. 
This procedure in fact removes high-$p_{\rm T}$ fake tracks, which originate from 
spurious matches of low $\pt$ particles in the TPC to hits in the ITS, and would result in an incorrect momentum assignment. 

Finally, tracks are rejected from the sample if their distance of
closest approach to the reconstructed vertex in the longitudinal
direction $d_{\rm z}$ is larger than 2 cm or $d_{\rm{xy}} > 0.018\,{\rm
cm}+0.035\,{\rm cm}\cdot p_{\rm T} ^{-1} $ in the transverse
direction with $p_{\rm T} $ in GeV/$c$,  which corresponds to 7 standard deviations of the resolution in $d_{\rm{xy}}$ (see  \cite{CHARM7TEV} for details).
The upper limit on the $d_{\rm z}$ ($d_{\rm z}<2$~cm) was set to minimize the contribution of tracks coming from pileup and beam-gas background events. These cuts reject less than 0.5\% of the reconstructed tracks independently of $\pt$ and collision centrality.

The efficiency and purity of the primary charged particle selection are estimated using a Monte Carlo simulation with HIJING~\cite{HIJING} 
events and a GEANT3~\cite{GEANT3} model of the detector response. 
We used a HIJING tune which reproduces approximately the measured
charged particle density in central collisions~\cite{ALICE_MULT}.
In the most central events, the overall primary charged particle 
reconstruction efficiency (tracking efficiency and acceptance) in $|\eta|<0.8$ is 36\% at $p_{\rm T} =0.15$~GeV/$c$
and increases to 65\% for $p_{\rm T}>0.6$~GeV/$c$.
In the most peripheral events the efficiency is larger than that for
the central events by about 1--3\%.
The contribution from secondary particles was estimated using the $d_{\rm{xy}}$ distributions of 
data and HIJING and is consistent with the measured strangeness to charged particle ratio from the reconstruction 
of K$^0_{\rm{s}}$, $\Lambda$ and $\overline{\Lambda}$ invariant mass peaks in \mbox{Pb--Pb} \cite{QM11Floris}. 
The total contribution from secondary tracks at $p_{\rm T} =0.15$~GeV/$c$
is 13 (7)\% for central (peripheral) events and decreases to about
0.6\% above $p_{\rm T} =4$~GeV/$c$ for both central and peripheral events.
From a systematic variation of the $\chi_{\rm{TPC-ITS}}^2$ cut and comparison of track properties in MC to data we conclude that the number of properly reconstructed tracks rejected as high-$\pt$ fake tracks is around 1--2\% (0.5--1\%) in the most central (peripheral) collisions. We also conclude that the contribution from the high-$\pt$ fake tracks to the $\pt$ spectra is negligible independently of the collision centrality and $\pt$.

\begin{table}
\centering
\caption{\label{SYSTERRORS} Contribution to the systematic uncertainties on the $p_{\rm T}$ spectra (0.15--50 GeV/$c$) for the most central and peripheral Pb--Pb collisions. Also listed are the systematic uncertainties on the pp reference (0.15--50 GeV/$c$) \cite{PPREFERENCE}.}
\begin{tabular}{|l|l|l|l|l|}
\hline
Centrality class & 0--5\%  & 70--80\%   \\
\hline
Centrality selection & 0.4\% & 6.7\%   \\
Event selection & 3.2\% & 3.4\% \\
Track selection & 4.1--7.3\% & 3.6--6.0\% \\
Tracking efficiency & 5\% &  5\% \\
$p_{\rm T}$ resolution correction & $<$1.8\%  & $<$3\% \\
Material budget & 0.9--1.2\% & 0.5--1.7\% \\
Particle composition & 0.6--10\% & 0.5--7.7\% \\
MC generator & 2.5\% & 1.5\% \\
Secondary particle rejection & $<$1\% & $<$1\%  \\
\hline
Total for $p_{\rm T} $ spectra &  8.2--13.5\% & 10.3--13.4\%  \\
\hline
\hline
Total for pp reference & \multicolumn{2}{|l|}{6.3--18.8\%} \\
pp reference normalization & \multicolumn{2}{|l|}{1.9\%} \\
\hline
\end{tabular}
\end{table}

The transverse momentum of charged particles is reconstructed from the track
curvature measured in the magnetic field $B=0.5$~T using the ITS and TPC detectors.  The $p_{\rm T}$ resolution is estimated
from the track residuals to the momentum fit and verified by cosmic 
muon events, and the width of the invariant mass peaks of $\Lambda$, $\overline{\Lambda}$ and K$^0_{\rm{s}}$ reconstructed from their decays to two charged particles. For the selected tracks the relative $p_{\rm T}$ resolution  ($\sigma(p_{\rm T})/p_{\rm T}$) amounts to 3.5\% at $p_{\rm T}=0.15$~GeV/$c$, has a minimum of 1\% at  $p_{\rm T}=1$~GeV/$c$, and increases linearly  to 10\% at  $p_{\rm T}=50$~GeV/$c$. It is independent of the centrality of the selected events. From the study of the invariant mass distributions of $\Lambda$ and K$^0_{\rm{s}}$ as a function of $p_{\rm T}$ we estimate that the relative uncertainty on the $p_{\rm T}$ resolution is around 20\%. From the mass difference between $\Lambda$ and $\overline{\Lambda}$ and the
ratio of positively to negatively charged tracks, assuming charge
symmetry at high $p_{\rm T}$, the upper limit of the systematic uncertainty of the momentum scale 
is estimated to be $|\Delta(p_{\rm T} )/p_{\rm T} |<0.005$ at $p_{\rm T}=50$~GeV/$c$. This has an 
effect of around 1.5\% on the yield of the measured spectra at the highest $p_{\rm T}$.
To account for the finite $p_{\rm T}$ resolution, correction factors for the reconstructed $p_{\rm T}$ spectra at $p_{\rm T} > 10$~GeV/$c$
 are derived using a folding procedure. 
The corrections depend on collision centrality due to the change
 of the spectral shape and reach 4 (8)\% at $p_{\rm {T}} = 50$~GeV/$c$ in the most central (peripheral) collisions.


\begin{figure}[t]
\begin{center}
\includegraphics[width= 3.6in,height=4.6in]{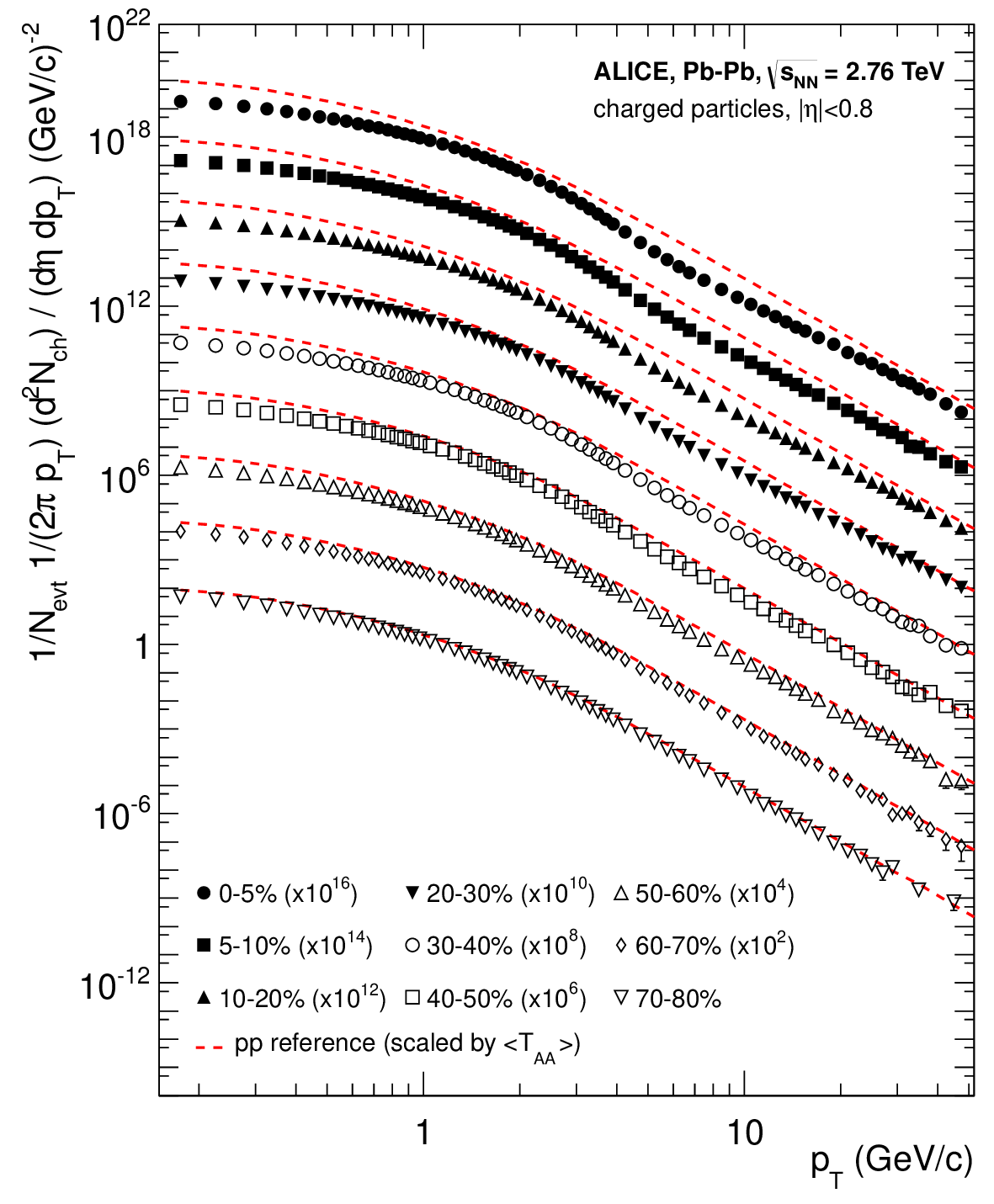}
\end{center}
\caption{\label{PbPbPTSPECRA} Charged particle $p_{\rm T}$ distribution measured in \mbox{Pb--Pb} collisions in different centrality intervals. The spectra are scaled for better visibility.  The dashed lines show the pp reference \cite{PPREFERENCE} spectra scaled by the nuclear overlap function determined for each centrality interval (Table~\ref{CENTR}) and by the Pb-Pb spectra scaling factors. The systematic and statistical uncertainties for \mbox{Pb--Pb} are added quadratically. The uncertainties on the pp reference are not shown.}
\end{figure}

The systematic uncertainties on the $p_{\rm T}$ spectra are summarized in Table \ref{SYSTERRORS}. 
The systematic uncertainties related to centrality selection were estimated by a comparison of the $p_{\rm{T}}$ spectra when the limits of the  centrality classes are shifted 
by $\pm1\%$ (e.g. for the 70--80\% centrality class, 70.7--80.8\% and 69.3--79.2\%), which is a relative uncertainty on the fraction of the hadronic cross section used in the Glauber fit \cite{ALICE_MULT} to determine the centrality classes. 
We also varied the event and track quality selection criteria and the Monte Carlo assumptions to estimate systematic uncertainties on the $\pt$ spectra.
In particular, we studied a variation of the most abundant charged
particle species (pions, kaons, protons) by $\pm30$\% to match the measured ratios and their uncertainties \cite{QM11Floris}. The material budget was varied
by $\pm$7\% \cite{KochBock}, and the secondary yield from strangeness decays in the
Monte Carlo by $\pm 30$\% to match the measured $d_{\rm{xy}}$ distributions. Moreover, we used a different event generator, DPMJET \cite{DPMJET}, to calculate MC correction maps.
The systematic uncertainties on the $p_{\rm T}$ spectra, related to the high-$p_{\rm{T}}$ fake track rejection procedure, were estimated
by varying the track matching criteria in the range $25<\chi_{\rm{TPC-ITS}}^{2}<49$, and amount to 1--4\% (1--2\%) 
in the most central (peripheral)  collisions. 
The total systematic uncertainties on the corrected $p_{\rm T} $
spectra depend on $p_{\rm T} $ and event centrality and amount to 8.2--13.5\% (10.3--13.4\%)  in the most central (peripheral)  collisions.

\begin{figure}[t]
\begin{center}
\includegraphics[width=4.5in]{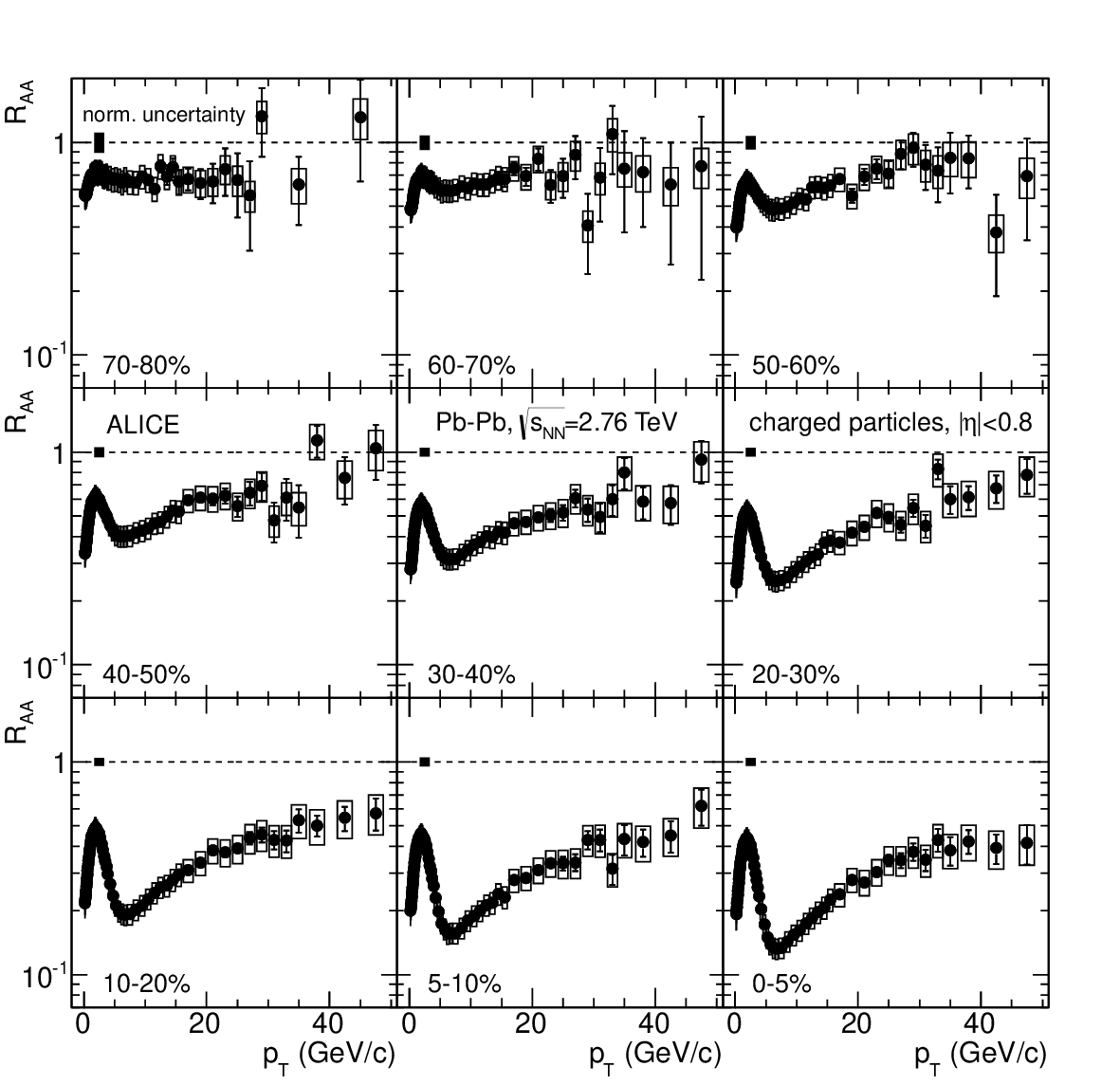}
\end{center}
\caption{\label{RAASPECTRA} Nuclear modification factor $R_{\rm{AA}} $ of charged particles measured in \mbox{Pb--Pb} collisions in nine centrality intervals. The boxes around data points denote $p_{{\rm T}}$-dependent systematic uncertainties. The systematic uncertainties on the normalization which are related to $\langle T_{\rm{AA}} \rangle$ and the normalization of the pp data are added in quadrature and shown as boxes at $R_{\rm{AA}} =1$. }
\end{figure}

A dedicated run of the LHC to collect pp reference data at
$\sqrt{s}=2.76$~TeV took place in March 2011. Data taken in this run were used to measure the charged particle $p_{\rm T}$  spectrum that forms the basis of the pp reference spectrum for $R_{\rm{AA}}$. Using these data the
systematic uncertainties in $R_{\rm {AA}}$ related to the pp reference
could be significantly improved (Table \ref{SYSTERRORS}) compared to the previous publication \cite{ALICE_RAA}, allowing for an exploration of
high-${p_{\rm T}}$ particle suppression in \mbox{Pb--Pb} out to 50
GeV/$c$. More details about the pp reference determination can be found in \cite{PPREFERENCE}. 


\section{Results}

The fully corrected $p_{\rm T}$ spectra of inclusive charged particles measured in \mbox{Pb--Pb} collisions at $\sqrt{s_{\rm{NN}}} = 2.76$~TeV in nine different centrality intervals, and the scaled pp reference spectra are shown in Fig.~\ref{PbPbPTSPECRA}. 
At low $\pt$, the transverse momentum spectra differ from the pp reference. This is in agreement with the previously observed scaling behavior of the total charged particle production as a function of centrality  \cite{ALICE_MULT}. A marked depletion of the spectra at high transverse momentum ($p_{\rm T} >5$~GeV/c) develops gradually as centrality increases, indicating strong suppression of high-$p_{{\rm T}}$ particle production in central collisions. 

The nuclear modification factors for nine centrality intervals are shown in Fig.~\ref{RAASPECTRA}. 
In peripheral collisions (70--80\%), only moderate suppression ($R_{\rm{AA}}=0.6$--0.7) and a weak $\pt$ dependence is observed.
Towards more central collisions, a pronounced minimum at about $p_{\rm T}=6$--7~GeV/$c$ develops while for \mbox{$p_{\rm T}>7$~GeV/$c$} there is a significant rise of the nuclear modification factor.
This rise becomes gradually less steep with increasing $p_{\rm T}$.
In the most central collisions (0--5\%), the yield is most
suppressed, $R_{\rm{AA}} \approx0.13$ at $p_{\rm T}=6$--7~GeV/$c$, and $R_{\rm{AA}}$ reaches $\approx$~0.4 with no significant $\pt$ dependence for  $p_{\rm T}>30$~GeV/$c$.

\begin{figure}[!t]
\begin{center}
\includegraphics[width=3.5in]{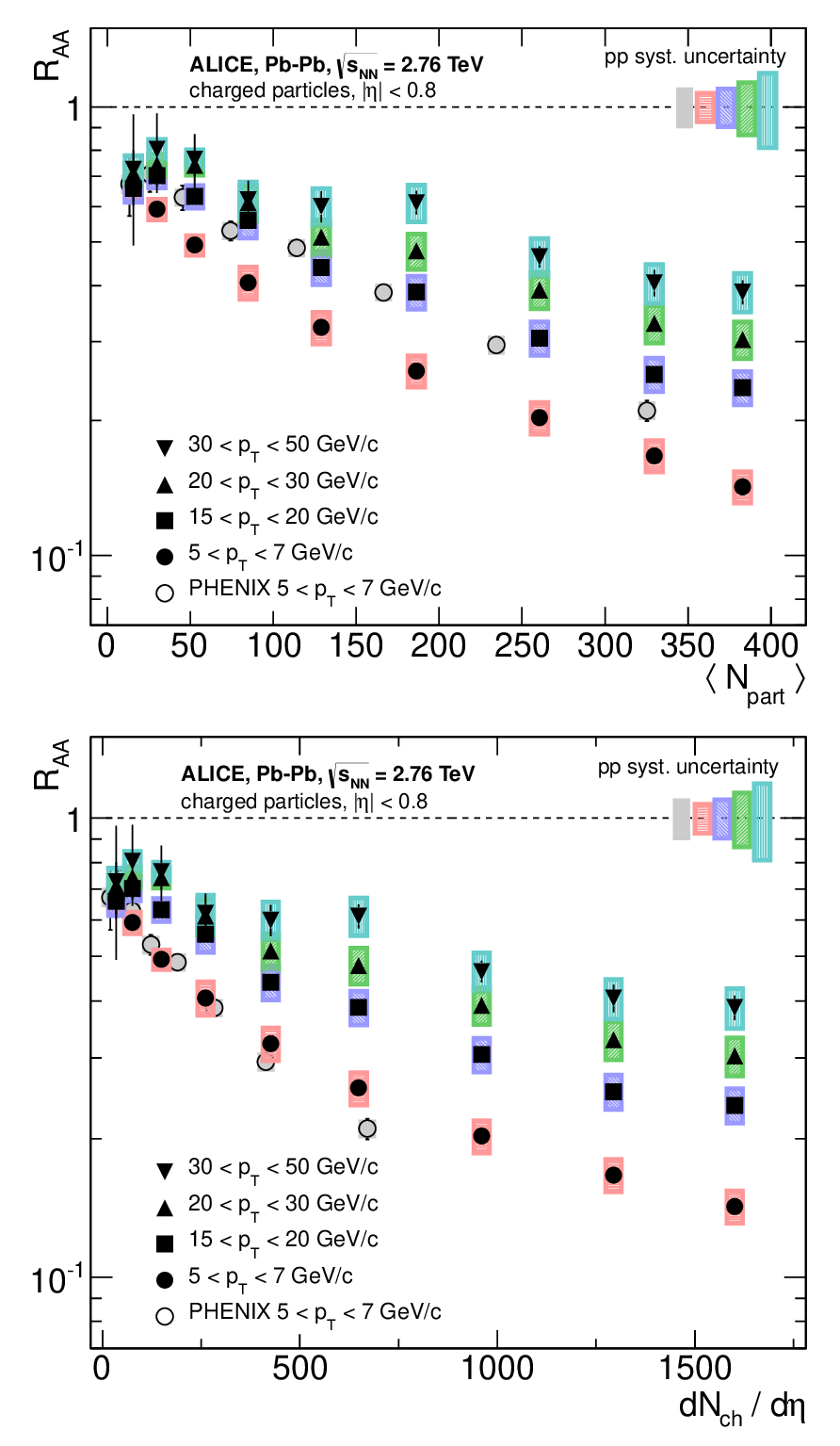}
\end{center}
\caption{\label{RAADNDETA} Nuclear modification factor $R_{\rm{AA}} $ of charged particles as a function of $\langle N_{{\rm{part}}}
 \rangle$ (top panel) and ${\rm{d}}{N_{\rm{ch}}}/\rm{d}\eta$ (bottom panel) measured by ALICE in \mbox{Pb--Pb} collisions in
different $\pt$-intervals, compared to PHENIX results in $5<\pt<7$
GeV/$c$ \cite{PHENIXRAA}. The boxes around the data represent
the $p_{\rm T}$-dependent uncertainties on the \mbox{Pb--Pb} $p_{\rm T}$ spectra. The boxes at $R_{\rm{AA}} =1$ represent the systematic uncertainties on the pp reference in different $\pt$-intervals ($p_{\rm{T}}$-interval increases from left to right, the left-most is for PHENIX). The systematic
uncertainties on the overall normalization for ALICE and PHENIX are not shown.}
\end{figure}

The dependence of $R_{\rm {AA}}$ on the collision centrality, expressed in terms of $N_{\rm {part}}$ and the charged particle multiplicity density (${\rm{d}}N_{\rm{ch}}/\rm{d}{\eta}$),
are shown in Fig.~\ref{RAADNDETA} for different intervals of $p_{\rm{T}}$. Also shown are results from PHENIX at RHIC 
in \mbox{Au--Au} collisions at \mbox{$\sqrt{s_{\rm{NN}}}=200$~GeV}  \cite{PHENIXRAA}. The
strongest centrality dependence is observed for particles with $5<p_{\rm T}<7$~GeV/$c$. At higher $p_{\rm T}$, the centrality dependence
weakens gradually.
In comparison to results from RHIC, the LHC data in the same 
$p_{\rm T}$ window show
a suppression which is larger by a factor of about 1.2 at all
 $\langle N_{\rm{part}} \rangle$ (Fig. \ref{RAADNDETA}, top panel). This implies that the shape of the  $N_{\rm {part}}$
dependence at RHIC and the LHC is very similar when the same $p_{\rm T}$ is 
compared, indicating a strong relation between collision geometry and energy loss.
The overall increase of suppression at the LHC as compared to RHIC may 
be expected from the larger density and longer lifetime of the fireball.
The suppression reaches similar values when results from
RHIC are compared to results from the LHC 
in terms of  ${\rm{d}}N_{\rm{ch}}/\rm{d}{\eta}$, 
as shown in Fig.  \ref{RAADNDETA} (bottom panel).
Larger values of suppression than at RHIC are observed in central collisions
at the LHC, where the charged particle multiplicity exceeds that of the
most central collisions at RHIC. It should be noted that the suppression at a given centrality results from a subtle interplay between the parton $\pt$ spectrum, the quark-to-gluon ratio, and the medium density, all of which exhibit a significant energy
dependence. Further model studies are needed to evaluate their relative contributions.

\begin{figure}[t]
\begin{center}
\includegraphics[width=4.5in]{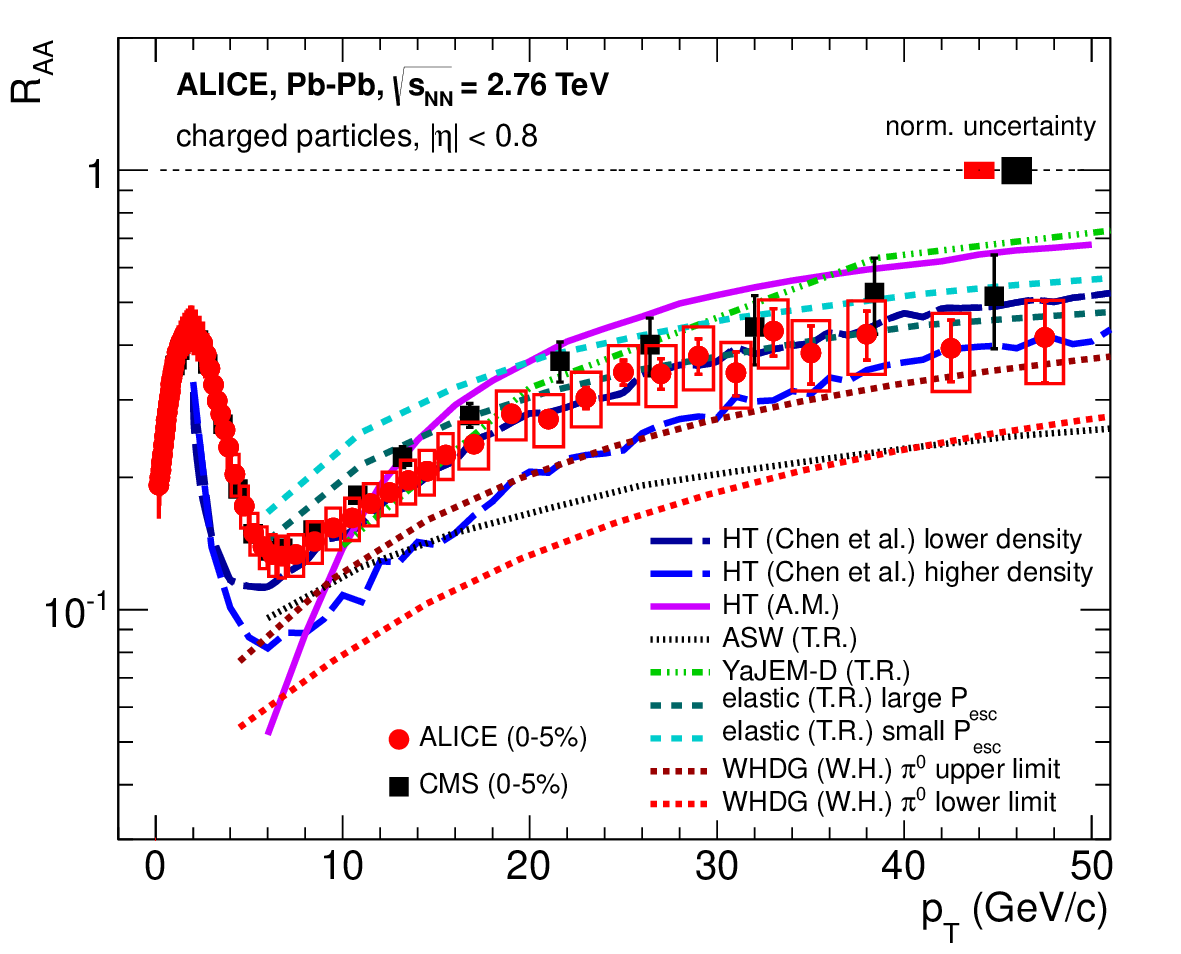}
\end{center}
\caption{\label{RAAMODELS} Nuclear modification factor $R_{\rm{AA}} $ of charged particles
 measured by ALICE in the most central \mbox{Pb--Pb} collisions (0--5\%)
 in comparison to results from CMS \cite{CMSRAA} and model calculations \cite{HOROWITZ2011, ASW, CHEN2011, MAJUMDER2011, RENK2011, RENK2011YAJEM}. The boxes around the data
 denote $p_{{\rm T}}$-dependent systematic
 uncertainties. For CMS statistical and systematic uncertainties on $R_{\rm {AA}}$ are added in quadrature. The systematic uncertainties on the normalization which are related to $\langle T_{\rm {AA}} \rangle$ and the normalization of the pp data are added in quadrature and shown as boxes at $R_{\rm{AA}} =1$ (the right-most is for CMS).}
\end{figure}

The ALICE measurement of $R_{\rm{AA}}$ in the most central \mbox{Pb--Pb} collisions (0--5\%) is compared to the CMS result \cite{CMSRAA} in Fig. \ref{RAAMODELS}.  Both measurements agree within their respective statistical and systematic uncertainties.

In Fig. \ref{RAAMODELS}, the measured $R_{\rm{AA}}$ for 0--5\% central collisions is also
compared to model calculations. All selected models use RHIC data to calibrate the medium density and were available before the preliminary version of the data reported in this paper. 
All model calculations except WHDG \cite{HOROWITZ2011} use a hydrodynamical description of the
medium, but different extrapolation assumptions from RHIC to
LHC. A variety of energy loss formalisms is used. An increase of $R_{\rm{AA}}$ due to a decrease of the relative
energy loss with increasing $p_{\rm T}$  is seen for all the models. 

The curves labeled WHDG, ASW, and Higher Twist (HT) are based on analytical radiative
energy loss formulations that include interference effects. Of those
curves, the multiple soft gluon approximation (ASW \cite{ASW}) and the
opacity expansion (WHDG \cite{HOROWITZ2011}) show a larger suppression
than seen in the measurement, while one of the HT curves
(Chen \cite{CHEN2011}) with lower density provides a good description. The other HT
(Majumder \cite{MAJUMDER2011}) curve shows a stronger rise with
$p_{\rm T}$ than measured. The elastic energy loss model by Renk (elastic)
\cite{RENK2011} does not rise steeply enough with $p_{\rm T}$
and overshoots the data at low $p_{\rm T}$. The YaJEM-D model \cite{RENK2011YAJEM}, which is
based on medium-induced virtuality increases in a parton shower, shows
too strong a $p_{\rm T}$-dependence of $R_{\rm{AA}}$ due to
a formation time cut-off.  

A more systematic study of the energy loss
formalisms, preferably with the same model(s) for the medium density
is needed to rule out or confirm the various effects.  Deviations of
the nuclear parton distribution functions (PDFs) from a simple scaling
of the nucleon PDF with mass number $A$ (e.g. shadowing) are also
expected to affect the nuclear modification factor. These effects are
predicted to be small for $p_{\rm T}>10$~GeV/$c$ at the LHC  \cite{HOROWITZ2011}  and
will be quantified in future p--Pb measurements.

\section{Summary}

We have reported the measurements of charged particle $p_{\rm T}$ spectra and nuclear modification factors $R_{\rm{AA}} $ as a
function of event centrality in \mbox{Pb--Pb} collisions at
$\sqrt{s_{\rm{NN}}} = 2.76$~TeV.  The results indicate a strong
suppression of charged particle production in \mbox{Pb--Pb} collisions and a
characteristic centrality and $p_{\rm T}$ dependence of the nuclear
modification factors.  In central collisions (0--5\%) the yield is most
strongly suppressed ($R_{\rm{AA}} \approx0.13$) at $p_{\rm T}=6$--7~GeV/$c$. Above $p_{\rm T}=7$~GeV/$c$, there is a significant rise
in the nuclear modification factor, which reaches $R_{\rm{AA}} \approx$~
0.4 for $p_{\rm T}>30$~GeV/$c$. This result is in agreement with the CMS measurement within statistical and systematic uncertainties. The suppression is weaker in peripheral
collisions  (70--80\%) with $R_{\rm{AA}}=0.6$--0.7 and no strong $\pt$ dependence. The observed suppression of high-$p_{\rm T} $ particles in central \mbox{Pb--Pb} collisions provides evidence
for strong parton energy loss and a large medium density at the LHC.
We observe that the suppression of charged particles with $5<p_{\rm T}<7$~GeV/$c$ reaches similar values when results from
RHIC are compared to results from LHC 
in terms of the ${\rm{d}}N_{\rm{ch}}/\rm{d}{\eta}$. The measured $R_{\rm{AA}}$ in \mbox{0--5\%} central collisions is compared to model calculations. An increase of $R_{\rm{AA}}$ due to a decrease of the relative
energy loss with increasing $p_{\rm T}$ is seen for all the models. The measurement presented here, together with measurements of particle
correlations \cite{ALICE_IAA} and measurements using jet
reconstruction \cite{ALICE_JETS}, will help in understanding
the mechanism of jet quenching and the properties of the medium
produced in heavy-ion collisions.
%
\newenvironment{acknowledgement}{\relax}{\relax}
\begin{acknowledgement}
\section{Acknowledgements}
The ALICE collaboration would like to thank all its engineers and technicians for their invaluable contributions to the construction of the experiment and the CERN accelerator teams for the outstanding performance of the LHC complex.
\\
The ALICE collaboration acknowledges the following funding agencies for their support in building and
running the ALICE detector:
 \\
Calouste Gulbenkian Foundation from Lisbon and Swiss Fonds Kidagan, Armenia;
 \\
Conselho Nacional de Desenvolvimento Cient\'{\i}fico e Tecnol\'{o}gico (CNPq), Financiadora de Estudos e Projetos (FINEP),
Funda\c{c}\~{a}o de Amparo \`{a} Pesquisa do Estado de S\~{a}o Paulo (FAPESP);
 \\
National Natural Science Foundation of China (NSFC), the Chinese Ministry of Education (CMOE)
and the Ministry of Science and Technology of China (MSTC);
 \\
Ministry of Education and Youth of the Czech Republic;
 \\
Danish Natural Science Research Council, the Carlsberg Foundation and the Danish National Research Foundation;
 \\
The European Research Council under the European Community's Seventh Framework Programme;
 \\
Helsinki Institute of Physics and the Academy of Finland;
 \\
French CNRS-IN2P3, the `Region Pays de Loire', `Region Alsace', `Region Auvergne' and CEA, France;
 \\
German BMBF and the Helmholtz Association;
\\
General Secretariat for Research and Technology, Ministry of
Development, Greece;
\\
Hungarian OTKA and National Office for Research and Technology (NKTH);
 \\
Department of Atomic Energy and Department of Science and Technology of the Government of India;
 \\
Istituto Nazionale di Fisica Nucleare (INFN) of Italy;
 \\
MEXT Grant-in-Aid for Specially Promoted Research, Ja\-pan;
 \\
Joint Institute for Nuclear Research, Dubna;
 \\
National Research Foundation of Korea (NRF);
 \\
CONACYT, DGAPA, M\'{e}xico, ALFA-EC and the HELEN Program (High-Energy physics \\ Latin-American--European Network);
 \\
Stichting voor Fundamenteel Onderzoek der Materie (FOM) and the Nederlandse Organisatie voor Wetenschappelijk Onderzoek (NWO), Netherlands;
 \\
Research Council of Norway (NFR);
 \\
Polish Ministry of Science and Higher Education;
 \\
National Authority for Scientific Research - NASR (Autoritatea Na\c{t}ional\u{a} pentru Cercetare \c{S}tiin\c{t}ific\u{a} - ANCS);
 \\
Federal Agency of Science of the Ministry of Education and Science of Russian Federation, International Science and
Technology Center, Russian Academy of Sciences, Russian Federal Agency of Atomic Energy, Russian Federal Agency for Science and Innovations and CERN-INTAS;
 \\
Ministry of Education of Slovakia;
 \\
Department of Science and Technology, South Africa;
 \\
CIEMAT, EELA, Ministerio de Educaci\'{o}n y Ciencia of Spain, Xunta de Galicia (Conseller\'{\i}a de Educaci\'{o}n),
CEA\-DEN, Cubaenerg\'{\i}a, Cuba, and IAEA (International Atomic Energy Agency);
 \\
Swedish Research Council (VR) and Knut $\&$ Alice Wallenberg
Foundation (KAW);
 \\
Ukraine Ministry of Education and Science;
 \\
United Kingdom Science and Technology Facilities Council (STFC);
 \\
The United States Department of Energy, the United States National
Science Foundation, the State of Texas, and the State of Ohio.
\end{acknowledgement}
%

%
\newpage
%
%
\appendix
\section{The ALICE Collaboration}
\label{app:collab}

\begingroup
\small
\begin{flushleft}
B.~Abelev\Irefn{org1234}\And
J.~Adam\Irefn{org1274}\And
D.~Adamov\'{a}\Irefn{org1283}\And
A.M.~Adare\Irefn{org1260}\And
M.M.~Aggarwal\Irefn{org1157}\And
G.~Aglieri~Rinella\Irefn{org1192}\And
A.G.~Agocs\Irefn{org1143}\And
A.~Agostinelli\Irefn{org1132}\And
S.~Aguilar~Salazar\Irefn{org1247}\And
Z.~Ahammed\Irefn{org1225}
% ...
\renewcommand\labelenumi{\textsuperscript{\theenumi}~}
\section*{Affiliation notes}
\renewcommand\theenumi{\roman{enumi}}
\begin{Authlist}
\item \Adef{M.V.Lomonosov Moscow State University, D.V.Skobeltsyn Institute of Nuclear Physics, Moscow, Russia}Also at: M.V.Lomonosov Moscow State University, D.V.Skobeltsyn Institute of Nuclear Physics, Moscow, Russia
\item \Adef{University of Belgrade, Faculty of Physics and Institute of Nuclear Sciences, Belgrade, Serbia}Also at: University of Belgrade, Faculty of Physics and "Vin\v{c}a" Institute of Nuclear Sciences, Belgrade, Serbia
\end{Authlist}
\section*{Collaboration Institutes}
\renewcommand\theenumi{\arabic{enumi}~}
\begin{Authlist}
\item \Idef{org1279}Benem\'{e}rita Universidad Aut\'{o}noma de Puebla, Puebla, Mexico
\item \Idef{org1220}Bogolyubov Institute for Theoretical Physics, Kiev, Ukraine
\item \Idef{org1262}Budker Institute for Nuclear Physics, Novosibirsk, Russia
\item \Idef{org1292}California Polytechnic State University, San Luis Obispo, California, United States
\item \Idef{org1329}Central China Normal University, Wuhan, China
\item \Idef{org14939}Centre de Calcul de l'IN2P3, Villeurbanne, France
\item \Idef{org1197}Centro de Aplicaciones Tecnol\'{o}gicas y Desarrollo Nuclear (CEADEN), Havana, Cuba
\item \Idef{org1242}Centro de Investigaciones Energ\'{e}ticas Medioambientales y Tecnol\'{o}gicas (CIEMAT), Madrid, Spain
\item \Idef{org1244}Centro de Investigaci\'{o}n y de Estudios Avanzados (CINVESTAV), Mexico City and M\'{e}rida, Mexico
\item \Idef{org1335}Centro Fermi -- Centro Studi e Ricerche e Museo Storico della Fisica ``Enrico Fermi'', Rome, Italy
\item \Idef{org17347}Chicago State University, Chicago, United States
\item \Idef{org1288}Commissariat \`{a} l'Energie Atomique, IRFU, Saclay, France
\item \Idef{org1294}Departamento de F\'{\i}sica de Part\'{\i}culas and IGFAE, Universidad de Santiago de Compostela, Santiago de Compostela, Spain
\item \Idef{org1106}Department of Physics Aligarh Muslim University, Aligarh, India
\item \Idef{org1121}Department of Physics and Technology, University of Bergen, Bergen, Norway
\item \Idef{org1162}Department of Physics, Ohio State University, Columbus, Ohio, United States
\item \Idef{org1300}Department of Physics, Sejong University, Seoul, South Korea
\item \Idef{org1268}Department of Physics, University of Oslo, Oslo, Norway
\item \Idef{org1132}Dipartimento di Fisica dell'Universit\`{a} and Sezione INFN, Bologna, Italy
\item \Idef{org1270}Dipartimento di Fisica dell'Universit\`{a} and Sezione INFN, Padova, Italy
\item \Idef{org1315}Dipartimento di Fisica dell'Universit\`{a} and Sezione INFN, Trieste, Italy
\item \Idef{org1145}Dipartimento di Fisica dell'Universit\`{a} and Sezione INFN, Cagliari, Italy
\item \Idef{org1312}Dipartimento di Fisica dell'Universit\`{a} and Sezione INFN, Turin, Italy
\item \Idef{org1285}Dipartimento di Fisica dell'Universit\`{a} `La Sapienza' and Sezione INFN, Rome, Italy
\item \Idef{org1154}Dipartimento di Fisica e Astronomia dell'Universit\`{a} and Sezione INFN, Catania, Italy
\item \Idef{org1290}Dipartimento di Fisica `E.R.~Caianiello' dell'Universit\`{a} and Gruppo Collegato INFN, Salerno, Italy
\item \Idef{org1103}Dipartimento di Scienze e Innovazione Tecnologica dell'Universit\`{a} del Piemonte Orientale and Gruppo Collegato INFN, Alessandria, Italy
\item \Idef{org1114}Dipartimento Interateneo di Fisica `M.~Merlin' and Sezione INFN, Bari, Italy
\item \Idef{org1237}Division of Experimental High Energy Physics, University of Lund, Lund, Sweden
\item \Idef{org1192}European Organization for Nuclear Research (CERN), Geneva, Switzerland
\item \Idef{org1227}Fachhochschule K\"{o}ln, K\"{o}ln, Germany
\item \Idef{org1122}Faculty of Engineering, Bergen University College, Bergen, Norway
\item \Idef{org1136}Faculty of Mathematics, Physics and Informatics, Comenius University, Bratislava, Slovakia
\item \Idef{org1274}Faculty of Nuclear Sciences and Physical Engineering, Czech Technical University in Prague, Prague, Czech Republic
\item \Idef{org1229}Faculty of Science, P.J.~\v{S}af\'{a}rik University, Ko\v{s}ice, Slovakia
\item \Idef{org1184}Frankfurt Institute for Advanced Studies, Johann Wolfgang Goethe-Universit\"{a}t Frankfurt, Frankfurt, Germany
\item \Idef{org1215}Gangneung-Wonju National University, Gangneung, South Korea
\item \Idef{org1212}Helsinki Institute of Physics (HIP) and University of Jyv\"{a}skyl\"{a}, Jyv\"{a}skyl\"{a}, Finland
\item \Idef{org1203}Hiroshima University, Hiroshima, Japan
\item \Idef{org1254}Indian Institute of Technology Bombay (IIT), Mumbai, India
\item \Idef{org36378}Indian Institute of Technology Indore (IIT), Indore, India
\item \Idef{org1266}Institut de Physique Nucl\'{e}aire d'Orsay (IPNO), Universit\'{e} Paris-Sud, CNRS-IN2P3, Orsay, France
\item \Idef{org1277}Institute for High Energy Physics, Protvino, Russia
\item \Idef{org1249}Institute for Nuclear Research, Academy of Sciences, Moscow, Russia
\item \Idef{org1320}Nikhef, National Institute for Subatomic Physics and Institute for Subatomic Physics of Utrecht University, Utrecht, Netherlands
\item \Idef{org1250}Institute for Theoretical and Experimental Physics, Moscow, Russia
\item \Idef{org1230}Institute of Experimental Physics, Slovak Academy of Sciences, Ko\v{s}ice, Slovakia
\item \Idef{org1127}Institute of Physics, Bhubaneswar, India
\item \Idef{org1275}Institute of Physics, Academy of Sciences of the Czech Republic, Prague, Czech Republic
\item \Idef{org1139}Institute of Space Sciences (ISS), Bucharest, Romania
\item \Idef{org27399}Institut f\"{u}r Informatik, Johann Wolfgang Goethe-Universit\"{a}t Frankfurt, Frankfurt, Germany
\item \Idef{org1185}Institut f\"{u}r Kernphysik, Johann Wolfgang Goethe-Universit\"{a}t Frankfurt, Frankfurt, Germany
\item \Idef{org1177}Institut f\"{u}r Kernphysik, Technische Universit\"{a}t Darmstadt, Darmstadt, Germany
\item \Idef{org1256}Institut f\"{u}r Kernphysik, Westf\"{a}lische Wilhelms-Universit\"{a}t M\"{u}nster, M\"{u}nster, Germany
\item \Idef{org1246}Instituto de Ciencias Nucleares, Universidad Nacional Aut\'{o}noma de M\'{e}xico, Mexico City, Mexico
\item \Idef{org1247}Instituto de F\'{\i}sica, Universidad Nacional Aut\'{o}noma de M\'{e}xico, Mexico City, Mexico
\item \Idef{org23333}Institut of Theoretical Physics, University of Wroclaw, Poland
\item \Idef{org1308}Institut Pluridisciplinaire Hubert Curien (IPHC), Universit\'{e} de Strasbourg, CNRS-IN2P3, Strasbourg, France
\item \Idef{org1182}Joint Institute for Nuclear Research (JINR), Dubna, Russia
\item \Idef{org1143}KFKI Research Institute for Particle and Nuclear Physics, Hungarian Academy of Sciences, Budapest, Hungary
\item \Idef{org1199}Kirchhoff-Institut f\"{u}r Physik, Ruprecht-Karls-Universit\"{a}t Heidelberg, Heidelberg, Germany
\item \Idef{org20954}Korea Institute of Science and Technology Information, Daejeon, South Korea
\item \Idef{org1160}Laboratoire de Physique Corpusculaire (LPC), Clermont Universit\'{e}, Universit\'{e} Blaise Pascal, CNRS--IN2P3, Clermont-Ferrand, France
\item \Idef{org1194}Laboratoire de Physique Subatomique et de Cosmologie (LPSC), Universit\'{e} Joseph Fourier, CNRS-IN2P3, Institut Polytechnique de Grenoble, Grenoble, France
\item \Idef{org1187}Laboratori Nazionali di Frascati, INFN, Frascati, Italy
\item \Idef{org1232}Laboratori Nazionali di Legnaro, INFN, Legnaro, Italy
\item \Idef{org1125}Lawrence Berkeley National Laboratory, Berkeley, California, United States
\item \Idef{org1234}Lawrence Livermore National Laboratory, Livermore, California, United States
\item \Idef{org1251}Moscow Engineering Physics Institute, Moscow, Russia
\item \Idef{org1140}National Institute for Physics and Nuclear Engineering, Bucharest, Romania
\item \Idef{org1165}Niels Bohr Institute, University of Copenhagen, Copenhagen, Denmark
\item \Idef{org1109}Nikhef, National Institute for Subatomic Physics, Amsterdam, Netherlands
\item \Idef{org1283}Nuclear Physics Institute, Academy of Sciences of the Czech Republic, \v{R}e\v{z} u Prahy, Czech Republic
\item \Idef{org1264}Oak Ridge National Laboratory, Oak Ridge, Tennessee, United States
\item \Idef{org1189}Petersburg Nuclear Physics Institute, Gatchina, Russia
\item \Idef{org1170}Physics Department, Creighton University, Omaha, Nebraska, United States
\item \Idef{org1157}Physics Department, Panjab University, Chandigarh, India
\item \Idef{org1112}Physics Department, University of Athens, Athens, Greece
\item \Idef{org1152}Physics Department, University of Cape Town, iThemba LABS, Cape Town, South Africa
\item \Idef{org1209}Physics Department, University of Jammu, Jammu, India
\item \Idef{org1207}Physics Department, University of Rajasthan, Jaipur, India
\item \Idef{org1200}Physikalisches Institut, Ruprecht-Karls-Universit\"{a}t Heidelberg, Heidelberg, Germany
\item \Idef{org1325}Purdue University, West Lafayette, Indiana, United States
\item \Idef{org1281}Pusan National University, Pusan, South Korea
\item \Idef{org1176}Research Division and ExtreMe Matter Institute EMMI, GSI Helmholtzzentrum f\"ur Schwerionenforschung, Darmstadt, Germany
\item \Idef{org1334}Rudjer Bo\v{s}kovi\'{c} Institute, Zagreb, Croatia
\item \Idef{org1298}Russian Federal Nuclear Center (VNIIEF), Sarov, Russia
\item \Idef{org1252}Russian Research Centre Kurchatov Institute, Moscow, Russia
\item \Idef{org1224}Saha Institute of Nuclear Physics, Kolkata, India
\item \Idef{org1130}School of Physics and Astronomy, University of Birmingham, Birmingham, United Kingdom
\item \Idef{org1338}Secci\'{o}n F\'{\i}sica, Departamento de Ciencias, Pontificia Universidad Cat\'{o}lica del Per\'{u}, Lima, Peru
\item \Idef{org1316}Sezione INFN, Trieste, Italy
\item \Idef{org1271}Sezione INFN, Padova, Italy
\item \Idef{org1313}Sezione INFN, Turin, Italy
\item \Idef{org1286}Sezione INFN, Rome, Italy
\item \Idef{org1146}Sezione INFN, Cagliari, Italy
\item \Idef{org1133}Sezione INFN, Bologna, Italy
\item \Idef{org1115}Sezione INFN, Bari, Italy
\item \Idef{org1155}Sezione INFN, Catania, Italy
\item \Idef{org1322}Soltan Institute for Nuclear Studies, Warsaw, Poland
\item \Idef{org36377}Nuclear Physics Group, STFC Daresbury Laboratory, Daresbury, United Kingdom
\item \Idef{org1258}SUBATECH, Ecole des Mines de Nantes, Universit\'{e} de Nantes, CNRS-IN2P3, Nantes, France
\item \Idef{org1304}Technical University of Split FESB, Split, Croatia
\item \Idef{org1168}The Henryk Niewodniczanski Institute of Nuclear Physics, Polish Academy of Sciences, Cracow, Poland
\item \Idef{org17361}The University of Texas at Austin, Physics Department, Austin, TX, United States
\item \Idef{org1173}Universidad Aut\'{o}noma de Sinaloa, Culiac\'{a}n, Mexico
\item \Idef{org1296}Universidade de S\~{a}o Paulo (USP), S\~{a}o Paulo, Brazil
\item \Idef{org1149}Universidade Estadual de Campinas (UNICAMP), Campinas, Brazil
\item \Idef{org1239}Universit\'{e} de Lyon, Universit\'{e} Lyon 1, CNRS/IN2P3, IPN-Lyon, Villeurbanne, France
\item \Idef{org1205}University of Houston, Houston, Texas, United States
\item \Idef{org20371}University of Technology and Austrian Academy of Sciences, Vienna, Austria
\item \Idef{org1222}University of Tennessee, Knoxville, Tennessee, United States
\item \Idef{org1310}University of Tokyo, Tokyo, Japan
\item \Idef{org1318}University of Tsukuba, Tsukuba, Japan
\item \Idef{org21360}Eberhard Karls Universit\"{a}t T\"{u}bingen, T\"{u}bingen, Germany
\item \Idef{org1225}Variable Energy Cyclotron Centre, Kolkata, India
\item \Idef{org1306}V.~Fock Institute for Physics, St. Petersburg State University, St. Petersburg, Russia
\item \Idef{org1323}Warsaw University of Technology, Warsaw, Poland
\item \Idef{org1179}Wayne State University, Detroit, Michigan, United States
\item \Idef{org1260}Yale University, New Haven, Connecticut, United States
\item \Idef{org1332}Yerevan Physics Institute, Yerevan, Armenia
\item \Idef{org15649}Yildiz Technical University, Istanbul, Turkey
\item \Idef{org1301}Yonsei University, Seoul, South Korea
\item \Idef{org1327}Zentrum f\"{u}r Technologietransfer und Telekommunikation (ZTT), Fachhochschule Worms, Worms, Germany
\end{Authlist}
\end{flushleft}
\endgroup

%
\end{document}